\documentclass[runningheads]{cl2emult}

\usepackage{makeidx}  % allows index generation
\usepackage{graphicx} % standard LaTeX graphics tool
\usepackage{subeqnar} % subnumbers individual equations
\usepackage{multicol} % used for the two-column index
\usepackage{cropmark} % cropmarks for pages without
\usepackage{eso}      % placeholder for figures
\makeindex            % used for the subject index
\begin{document}
\title*{Abundances in Sagittarius: present state\protect\newline 
and perspectives
for the use of VLT
}
\toctitle{Abundances in Sagittarius:
\protect\newline  present state and perspectives
for the use of VLT
}
% allows explicit linebreak for the table of content
%
%
\titlerunning{Abundances in Sagittarius}
% allows abbreviation of title, if the full title is too long
% to fit in the running head
%
\author{Piercarlo Bonifacio}
\authorrunning{P. Bonifacio}
% if there are more than two authors,
% please abbreviate author list for running head
%
%
\institute{Osservatorio Astronomico di Trieste, 
Via G.B. Tiepolo 11, 34131, Trieste, Italia
     }

\maketitle              % typesets the title of the contribution

\begin{abstract}

Sagittarius, the nearest external galaxy, will  be amenable
to detailed abundance studies with VLT and other 8m class telescopes.
Such data, in conjuction with the similar data of our own galaxy, will
allow a deeper understanding of chemical evolution.  
Our study conducted with
NTT shows the presence of metal-rich stars with radial velocities
compatible with Sagittarius membership  and of stars as metal-poor as 
$\rm [Fe/H]=-1.5$.  
In this talk I shall address the way in which
VLT instruments will allow to clarify this intricate situation.

%%%%%%%%%%%%%%%%%%%%%%%%%%%%%%%%%%%%%%%%%%%%%%%%%%%%%%%%%%%%%%%%%%%%%%%%%
%%%%%%%%%%%%%%%%%%%%%%%%%%%%%%%%%%%%%%%%%%%%%%%%%%%%%%%%%%%%%%%%%%%%%%%%%

%%%%%%%%%%%%%%%%%%%%%%%%%%%%%%%%%%%%%%%%%%%%%%%%%%%%%%%%%%%%%%%%%%%%%%%%%
%%%%%%%%%%%%%%%%%%%%%%%%%%%%%%%%%%%%%%%%%%%%%%%%%%%%%%%%%%%%%%%%%%%%%%%%%

\end{abstract}

\section{Introduction}

The Sagittarius dwarf galaxy 
discovered \cite{ib94,ib95}  as a comoving group
($V_{hel}\approx 140 {\rm kms^{-1}}$) with a velocity
dispersion as small as $2 \rm kms^{-1}$,
is
the nearest external galaxy and will  be amenable
to detailed abundance studies with VLT and other 8m class telescopes.
These shall be important for two distinct reasons:
on the one hand, we shall be able to study chemical
evolution in an environment different from our Galaxy; on the other hand,
if Sagittarius shows any chemical signature, this should allow us to
identify Sagittarius debris, currently populating the Galactic Halo.

The colour-magnitude diagram of Sgr
shows a wide Red Giant Branch (RGB), which is 
interpreted as evidence of a spread in metallicity of Sgr.
Marconi et al \cite{mara} conclude that the RGB of Sgr lies between
that of 47 Tuc ([Fe/H]$=-0.71$) and M2 ([Fe/H]$=-1.58$).
In the following I shall describe the work done with the ESO NTT
to determine spectroscopic metallicities of giants in Sgr and shall
address the potentiality of VLT on this issue.

\section{NTT Observations}

We selected, from the Marconi et al \cite{mara} sample, stars
on the RGB which ought to  display the spread in metallicity
present in Sgr. We used the ESO NTT telescope with the EMMI instrument
in Multi Object Spectrocopy (MOS) mode. The dispersing element was
grism \# 5 providing a resolution of about 1500.
We have obtained spectra for a total of 57 objects in field Sgr1 of
\cite{mara} on June 19th 1996.
In addition, we obtained long slit spectra of one 
of the stars observed with the MOS and of star HD 190287
on September 18th 1998.

\begin{figure}
\centering
\includegraphics[]{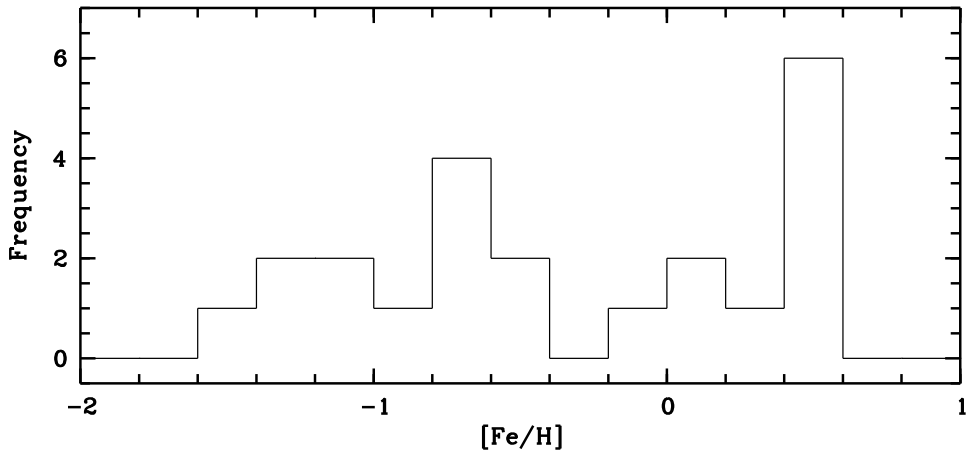}
\caption[]{Metallicity distribution for 22  radial velocity
members of Sgr }
\label{eps1}
\end{figure}

Of the 57 stars observed 23 matched the criterion
100 $\rm kms^{-1}\le v_{hel}\le 180 ~ kms^{-1}$ \cite{ib97}, which we adopted,
to ascribe membership to Sagittarius.
For these 23 stars the average radial velocity is 
 $136\rm ~kms^{-1}\pm 18 ~kms^{-1}$, 
in good agreement with previous determinations; the r.m.s.
is dominated by the error in the measure of the radial velocity 
and not by the velocity dispersion of Sgr.

\section{Abundance estimates from low resolution spectra}

Our  approach has
been to define spectral indices which measure some prominent spectral
features which may be used for abundance estimates. 
We defined six spectral indices, two of which measure the Mg I b triplet
and the remaining four measure essentially iron and iron-peak elements.
The location of the indices may be found in Fig.1 of reference\cite{marb} .
The indices are all measured with respect to a common pseudo-continuum,
defined by six quasi-continuum windows.
The aim is to be able to determine
both [Fe/H] and [Mg/Fe]. 

We computed the values of the indices for a small grid of synthetic
spectra (T=4750 K, 5000 K, 5250 K; log g = 2.50
[Fe/H]$=-1.5,-1.0,-0.5,0.0,+0.5$ ; $\rm \xi=2kms^{-1}$) computed
with the SYNTHE code \cite{Kur93}. 
The spectra where broadened with a gaussian profile
of 210 $\rm kms^{-1}$ to match the resolution of
the observed spectra.
The input model-atmospheres were
computed with version 9 of the ATLAS code\cite{Kur93} switching
off the overshooting option. Models with [Fe/H]$=-1.5$ and $-1.0$
were computed using $\alpha-$enhanced opacity distribution functions.
For each star
the temperature was determined from the $(V-I)_0$ colour\cite{mara} 
through the calibration of\cite{alo} .
Strictly speaking this calibration is valid only for dwarf stars,
however the $V-I$ colour has only a weak dependence on gravity,
especially in the colour range we are interested in.
We excluded from analysis SgrM 172 ($V-I=1.81$),  because it is much
cooler than all the other stars.
The metallicity was
determined from the four iron-sensitive 
indices  by interpolating in the table of computed indices. 

\subsection{Zero-Points}

We made the same analysis for the Sun using the solar atlas\cite{Kur84}, 
degraded at the same resolution as our
grism spectra, as an observed spectrum and a small
grid of synthetic spectra appropriate for the Sun
(T=5777, log g =4.44, [Fe/H]$=-0.5,0.0,+0.5$).
The difference between these derived metallicities and the
meteoritic 
iron abundance, is small
(around 0.1--0.2 dex).
On the assumption
that these differences reflect inadequacies of
our atomic data and model atmosphere and
that the systematics is the same for the Sun and for our program
stars, we treated these  as zero-point shifts. 

\subsection{Cross-checks}

We determined the metallicity of the Sun  from 
a twilight spectrum taken with the MOS. The deduced
[Fe/H] was $-0.32,-0.04,-0.01,+0.10$ for the four indices
respectively. These  
are not 0.00 because of   the noise in the 
observed spectrum. One
of the indices (F1)  yields an abundance considerably
different from the others. 
To obtain the final metallicity we decided
to use a central estimate, defined as the mean of the two central
values, which is akin to the median and more robust than the mean. In this
case the central estimate is [Fe/H]$=-0.02$ , in good agreement
with the expected value of 0.00.

Our program stars are considerably cooler and more luminous than the Sun.
We analysed the  spectrum of the cool metal-poor giant
HD 190287
([Fe/H]$=-1.34$, $\rm T_{eff}=5300$), 
taken with EMMI, grism \# 5 and a  long-slit.
We obtain
[Fe/H] $=-1.52,-1.22,-1.23,-1.37$, for the four indices, which
yields a central estimate [Fe/H]$=-1.30$.

We compared a spectrum of one of our
program stars (SgrM 139) through the EMMI long-slit with its MOS spectrum.
The derived metallicities were the same ($\rm[Fe/H]_{MOS}=-0.05$ ;
$\rm [Fe/H]_{LS} = +0.02$) and also the radial velocities
were consistent within errors ($v_{MOS}=+147 {\rm kms^{-1}}$ ;
$v_{LS}=+124 {\rm kms^{-1}}$).

\subsection{Error estimates}

In order to estimate the random errors in our metallicity estimates we
used a Montecarlo simulation. We took a synthetic spectrum
($\rm T_{eff} = 5000 K$
[Fe/H]$=-0.5$)
to which we added Poisson noise so that S/N = 10 (i.e. the S/N of our
spectra) and estimated the metallicity from this noisy spectrum.
The mean of 10000 realizations was $\rm <[Fe/H]>= -0.56$
and the standard deviation was 0.25 dex,
we  take this as our  estimate of the random error.

In addition one should consider the 
errors arising from uncertainties in the atmospheric parameters.
The largest error comes from the microturbulent velocity.
At this resolution we are forced to rely only on strong lines
which are very sensitive to microturbulence. The value
of $\xi$ must be assumed since there is no way to fix it from
the spectrum. The associated error is larger for the more metal-rich stars,
if the microturbulence is 1 $\rm kms^{-1}$ 
the metallicities
increase by $\rm \Delta([Fe/H]) = 0.24\times[Fe/H]+0.38 $.
A change in effective temperature by 150 K brings about
a change of 0.2 dex in metallicity as does a change of surface
gravity by 0.5 dex.

\subsection{Previous results and future work}

Preliminary results of this work have been already presented\cite{marb,bon}.
There are two
main differences with the results presented here.
The first is that here we use  pre-tabulated synthetic indices
rather than iteratively performing spectrum synthesis for
each star and each index. The second 
is that in the present work
the indices are  measured with respect to the pseudo-continuum,
rather than with respect to an estimate of the true continuum,  obtained
by iteratively re-normalizing the observed spectrum using the
current estimate of the best-matching synthetic spectrum, as done 
previously.
The re-normalization   procedure was abandoned because we 
realized that it is strongly
dependent on the initial metallicity estimate. In fact it tends
to relax the observed spectrum onto the first guess. Differences
on the order of 0.2 dex are obtained from the same observed spectrum
but initial metallicity guesses differing by 0.5 dex.
Our present results are in  agreement with our previous estimates
at the level of 0.1 dex except for the two most metal-poor stars
(SgrM 124 and SgrM 115)
for which we find a metallicity which is 0.32 and 0.41 dex lower.

We have so far not determined [Mg/Fe] 
because the indices M1 and M2 depend on both [Mg/H] and [Fe/H].
Our grid of synthetic spectra includes only ``standard'' values
of [Mg/Fe] (0.0 down to [Fe/H]$=-0.5$ and +0.4 below). On the other
hand, our preliminary results suggest  different [Mg/Fe] ratios,
especially for $\rm -0.5 < [Fe/H] <  0.0 $ .
On the contrary the F indices are essentially independent of
the abundance of $\alpha$ elements. Determining [Mg/Fe] is still
possible, once [Fe/H] is fixed through the F indices, but requires
a grid of synthetic spectra computed for a range of [Mg/Fe] values
or direct spectrum synthesis on a star by star basis. This 
shall be the object of our future work

\section{Results on Sagittarius}

The metallicity distribution of our 22 candidate Sagittarus members is
shown in Figure 1. The peak at metallicity +0.5 reflects our choice
to avoid extrapolation at metallicities 
higher than our most metal-rich grid point; when a star showed
indices indicative of higher metallicity we assigned to it the value +0.5.

The distribution appears to be distinctly tri-modal: there is a metal-rich
population
whose mean metallicity is yet undefined 
and there are two metal-poor populations
which peak around $-1.2$ and $-0.5$ respectively.
The existence of the two metal-poor populations is in substantial
agreement with previous findings\cite{mara}.
The presence of a metal-rich
population is something new and unexpected.
It is not yet clear whether the metal-rich stars belong to Sagittarius
or are Bulge interlopers, although in the latter case their
number is uncomfortably large and may require a revision of our
understanding of Bulge kinematics.
An independent study  based on Keck 
High-Res spectra\cite{sme}, finds that 2 out of 7 stars, with radial
velocity compatible with Sagittarius membership, 
are in fact metal-rich. Yet another photometric survey\cite{bel}
suggests the possible existence of a RGB sequence
considerably more metal-rich than 47 Tuc.

Although we are at an early stage to draw definitive conclusions,
circumstantial evidence is arising for a metal-rich
population in Sagittarius. This does not fit with
our present understanding of 
chemical evolution of dwarf spheroidals, however we should keep an open mind
and stick to the observations.

\section{Perspectives for the use of VLT}

VLT will quite likely play a fundamental role
in the study of Local Group Galaxies and Sagittarius in particular.
With UVES  it shall be possible 
to do a detailed line by line analysis
for stars along the RGB to below the Horizontal Branch
($V_{HB}\approx 18.2$). The Turn-off $21\le V \le 21.5$ will still
be out of reach of UVES though.

New perspectives are opened by FORS. The resolution shall be
at most 1/2 of that used  by us at NTT, however the indices
we have defined measure such strong features  that they will be usable
even at this lower resolution, as shown
by simulations we carried out. Moreover the MOS slitlets of FORS 
have sharp edges and  it will be possible
to flux-calibrate the spectra, thus  allowing to measure spectral
indices in absolute flux. 
This will recover all the information contained the continuum.
The spectral coverage (350 - 590 nm) will allow to use
the Ca II H and K lines and probably other iron-related indices
in the range 400 - 500 nm . Finally with VLT+FORS a S/N of about 100
can be reached in one hour for  stars of $V\approx 18.5$, with
this S/N our Montecarlo simulations predict
the random error in metallicity determinations to be around
0.02 dex. 

The use of a 3m class telescope, such as NTT, for
preliminary observations aimed at the determination of radial velocities
has proved to be too time-consuming to be worth the effort.
FORS will be able to provide good quality spectra from which 
radial velocities, metallicities and key abundance ratios, such as
[Mg/Fe] and [Ca/Fe], can be determined.

The real killer in the studies of Local Group galaxies shall
be Giraffe fed by 130 fibres in MEDUSA mode. The highest resolution
achieved by Giraffe (15000) is high enough to perform a classical
line by line analysis. However the number of spectra we shall get
(several hundreds per night) will preclude the use 
of this approach. We will have to find new ways to estimate
abundances from these spectra which will be  able to cope with
the data flow provided by Flames+Giraffe.

One possibility is suggested by our low-resolution experience: 
define spectral indices that   
can be related to abundances of particular elements and tabulate 
their value from synthetic spectra.
The estimation of an abundance requires then only an interpolation
in an appropriate table.

Different indices will be appropriate for different temperature and
luminosity regimes. Now is the time to investigate
both theoretically and observationally 
which features will provide useful indices.
For example some indices may be defined to measure essentially
strong iron lines others essentially weak iron lines,
the balance of the two will allow to determine microturbulence.

The index approach is not the only possibility. Other 
very promising techniques such
as autocorrelation and cross-correlation should be investigated.
A whole range of methods could complement each other to extract
information from the spectra, however if any particular
star seems deserving of special attention, all the spectra will be nicely
archived and one can go back and do the good old line by line analysis.

\section{Acknowledgements}

I wish to thank G. Marconi, P. Molaro and L. Pasquini for allowing
me to present results of common work in advance of publication.
This work is
based on observations collected at the European Southern Observatory, Chile,
ESO N$^\circ$ 57.E-0586.

%INDEX%%%%%%%%%%%%%%%%%%%%%%%%%%%%%%%%%%%%%%%%%%%%%%%%%%%%%%%%%%%%%%%
\clearpage
\addcontentsline{toc}{section}{Index}
\flushbottom
\printindex
%%%%%%%%%%%%%%%%%%%%%%%%%%%%%%%%%%%%%%%%%%%%%%%%%%%%%%%%%%%%%%%%%%%%%


\addcontentsline{toc}{section}{References}

\begin{thebibliography}{7}
\bibitem{alo}
Alonso A., Arribas S., Martinez-Roger C. (1996)
The empirical scale of temperatures of the low main sequence (F0V-K5V).
A\& A 313:873
\bibitem{bon} Bonifacio P., Pasquini L., Molaro P., Marconi G. (1998) 
Spectroscopy of giants of the Sagittarius dwarf galaxy,
In:Galaxy evolution:Connecting 
the distant Universe
       with the local fossil record Paris, Meudon 21-25 September 1998 
\bibitem{bel} Bellazini M., Ferraro F.R., Buonanno R. (1999)
The Sagiitarius Dwarf Galaxy Survey (SDGS) I  MNRAS, in press
astro-ph/9812344
\bibitem{ib94} Ibata R.A., Gilmore G., Irwin M.J. (1994)
A dwarf satellite galaxy in Sagittarius. Nature 370:194
\bibitem{ib95}Ibata R.A., Gilmore G., Irwin M.J. (1995)
                                   Sagittarius: the nearest dwarf galaxy.
MNRAS 277:781
\bibitem{ib97} Ibata R.A., Wyse R.F.G., Gilmore G., Irwin M.J.,
Suntzeff N.B. (1997) The Kinematics, Orbit, 
and Survival of the Sagittarius Dwarf Spheroidal Galaxy.
AJ 113:634
\bibitem{Kur93} Kurucz R.L. (1993) SAO CD-ROM No. 13,18
\bibitem{Kur84} Kurucz R.L., Furenlid I., Brault J., 
Testerman L. (1984) Solar Flux Atlas from 296 to 1300 nm.
Nat. Sol. Obs. Atlas No.1
\bibitem{mara}
Marconi G., Buonanno R., Castellani M., Iannicola G., Molaro P.,
Pasquini L., Pulone L. (1998a) The stellar content of the Sagittarius 
Dwarf Galaxy. A\& A 330:453
\bibitem{marb} Marconi G., Bonifacio P., Pasquini
P., Molaro P. Spectroscopy of
red giants of the Sagittarius dwarf galaxy (1998b)
In:
The Stellar Content of Local Group Galaxies, conference 
held in Cape Town, 
South Africa, 7-11 September 1998, to be published by ASP, edited by
                             Patricia Whitelock and Russell Cannon, p. 8.
\bibitem{sme}
Smecker-Hane T., McWilliam A., Ibata R.A. (1998)
Chemical Abundances in the Sagittarius Dwarf Galaxy. AAS 192:6613




\end{thebibliography}
\end{document}